\documentclass[prl, twocolumn, superscriptaddress, reprint]{revtex4-1}
\usepackage{amsmath}
\usepackage{mathtools} 
\usepackage{amsfonts}
\usepackage{physics}
\usepackage{amssymb}
\usepackage{graphicx}  
\usepackage{subfigure}
\usepackage{placeins}%
\usepackage{afterpage}%
\usepackage{hyperref}  
\usepackage{color}

\begin{document}
\bibliographystyle{apsrev4-1}

\title{Tuning the Delicate Topology of Topological Phases}
\author{Snigdh Sabharwal}
\email{snigdh.sabharwal@oist.jp}
\affiliation{Theory of Quantum Matter Unit, Okinawa Institute of Science and Technology Graduate University, Onna-son, Okinawa 904-0412, Japan}

\begin{abstract}
We present a unified framework to systematically embed complex knotted and linked structures, beyond the torus family, into diverse topological phases, including Hopf insulators, classical spin liquids, topological semimetals, and non-Hermitian metals. Using rational maps and level sets of complex polynomials, we explicitly construct new topological models exhibiting rich and previously inaccessible textures. These topological features manifest distinctly across physical systems: emergent magnetic field lines in Hopf insulators directly reflect the rational-map topology, paralleling topological electromagnetism, while in classical spin liquids the topology is experimentally accessible via the equal-time structure factor. Our approach thus provides both a conceptual unification of previously disconnected systems and a practical toolset for realizing and detecting intricate topological textures in experiments.
\end{abstract}

\maketitle

{\it Introduction.}
From an unsuccessful attempt to classify atoms based on different knot types \cite{Thomson1867ii,Knott1911life} to the classification of phases of matter using topological invariants  \cite{Ryu2010topological}, the applications of topology within physics boast a rich and varied history. The percolation of these ideas to various fields from plasma physics \cite{Moffatt2013topological} to quantum computing \cite{Kauffman1993quantum} highlights the general applicability of these methods. 

Within the field of topological phases of matter, one is driven by the search for new exotic phases. A powerful strategy in this pursuit is to use complex maps that encode topologically nontrivial information. A well-known example is the Hopf map \cite{Hopf1931abbildungen}, which was used to realize a three-dimensional topological insulator, the Hopf insulator \cite{Moore2008}. This approach was later generalized in \cite{Deng2013} by employing the Whitehead map \cite{Whitehead1947expression}. Similar ideas, utilizing other maps \cite{Brauner1928verhalten, Dennis2010isolated}, have been used to construct knotted and linked topological semimetals \cite{Ezawa2017,Chen2017,Yan2017} and non-Hermitian metals \cite{Carl2019}. Curiously, however, these constructions have thus far been confined to encoding torus knots and links. In contrast, complex maps that encode topological structures beyond the torus family have found great success in constructing approximate solutions of the Skyrme-Faddeev model \cite{Sutcliffe2007knots,Jennings2015cabling} and designing topologically nontrivial electromagnetic fields \cite{Kedia2016,Bode2017knotted,Arrayas2017knots}. These maps yield a remarkable diversity of knotting and linking structures, from torus knots and links to figure-8, cable knots, Borromean rings, etc., raising the question: can these complex topologies be effectively integrated into the framework of topological phases of matter?

In this work, we develop a unified methodology to construct complex polynomials that systematically incorporates complex topologies, extending beyond the torus class, into various topological phases of matter. By leveraging level sets of these polynomials, we construct models that capture a wide array of knotted and linked textures (such as figure‑8 knots, cable knots, etc.) and translate these into the language of topological insulators, semimetals, non-Hermitian metals and classical spin liquids (CSLs).
For instance, in the cases of Hopf insulators, and classical spin liquids, the level set topology is imprinted via complex rational maps, with the corresponding Hopf-Pontrjagin (HP) index \cite{Pontrjagin1941classification} serving as a topological marker. In Hopf insulators, this topology is directly reflected in the field line configuration of the emergent magnetic field. Additionally, in classical spin liquids, the level set topology is manifest onto the equal-time structure factor, which is experimentally accessible.  Furthermore, by leveraging the same polynomial construction, we readily transfer the encoded level set topology to $\mathcal{PT}$-symmetric semimetals and non-Hermitian metals, thereby generalizing earlier approaches. 

Our work provides systematic scheme to construct knotted and linked models of topological phases, highlighting the connections between different fields and paving out straightforward extensions to other models in the future. 
\begin{figure*}[ht!]
	\centering
	\includegraphics[page=1,scale=1.34]{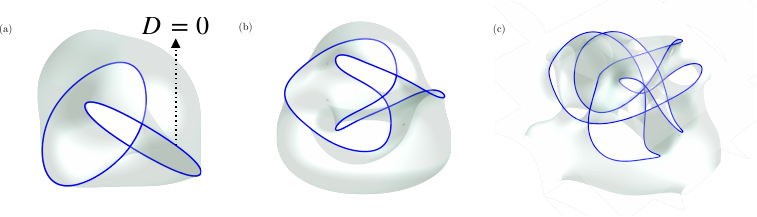}
	\caption{Visualizing the zero sets of complex polynomials, $D=0$ (in blue) . (a) The zero set of Eq.~(\ref{eq:Torusknotlinkmap}) for $\gamma = 2, \delta=2$, which corresponds to linked circles. 
		(b) The zero set of Eq.~(\ref{eq:fig8knot}), which corresponds to the figure-8 knot.
		(c) The zero set of Eq.~(\ref{eq:C2332cableknot}), which corresponds to the cable knot. The transparent surfaces are only meant to emphasize the knotted or linked structure. }
	\label{fig:ZeroSetD}
\end{figure*}

\begin{figure*}[ht!]
	\centering
	\includegraphics[page=2,scale=1.5]{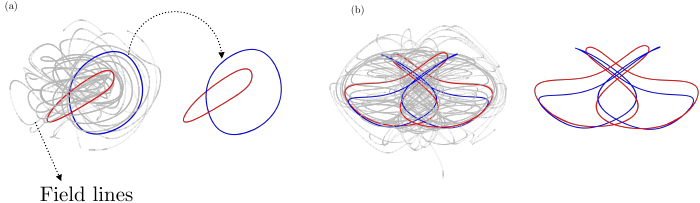}
	\caption{Observing the magnetic field lines (shown in gray) obtained via rational maps. (a) Field lines generated by the Hopf map [Eq.~(\ref{eq:Hopfmap})]. (b) Field lines generated by the Whitehead map [Eq.~(\ref{eq:Whiteheadmap}) with $p=2,\,q=3$]. The plots were obtained for $m=-2.3$. Red and blue lines highlight selected fibers, emphasizing the underlying level set topology of each map.}
	\label{fig:StreamlineRationalMap}
\end{figure*}

{\it Review of Hopf insulators\textbf{\----}}
Hopf insulators are three-dimensional TIs  featuring a single conduction and a valence band. These insulators are characterized by a non-vanishing topological invariant, the HP index \cite{Pontrjagin1941classification}. These insulators are modeled  as follows
\begin{equation}
	\mathcal{H}(\mathbf{k}) = \mathbf{\hat{n}(k)}  \cdot \boldsymbol{\sigma}\;, \quad | \mathbf{n(k)}| \neq  0  \;\; \forall \mathbf{k}\;,
\end{equation}
where $\boldsymbol{\sigma}$ are the Pauli matrices and $\mathbf{\hat{n}(k)} $  is the pseudo-spin field defined as 
\begin{equation}\label{eq:pseudospin}
	\mathbf{\hat{n}(k)} = \mathbf{\hat{z}(k)}^{\dagger}\boldsymbol{\sigma}\mathbf{\hat{z}(k)}\;.
\end{equation}
The topological texture of these insulators is imparted using,
\begin{equation}
	\mathbf{\hat{z}} = (z_{1},z_{2})^{T} \;, |z_1|^2 + |z_2|^2 = 1\;,
\end{equation}
where $z_{1}$ and $z_{2}$ are functions of 
\begin{align}
	u_{1} &= \sin(k_{x}) + i \sin(k_{y}) \\
	u_{2} &= \sin(k_{z}) + i \bigg(\sum\nolimits_{l = x,y,z} \cos(k_{l}) + m  \bigg)\;,
\end{align}
and their complex conjugates. Here, the parameter $m$ controls whether the topology is trivial or otherwise.

The characterization of the nontrivial topology is established by an invariant, a Hopf-Pontrjagin (HP) index \cite{Pontrjagin1941classification}, 
\begin{align}
	\chi &= \frac{1}{4\pi^{2}} \int_{BZ}d^{3}\mathbf{k}\; \mathbf{B}\cdot\mathbf{A} \;,
\end{align}
where,
\begin{align}\label{eq:AandB}
	\mathbf{A} &= i \bigg(z_{1}^{*}\grad z_{1} +z_{2}^{*}\grad z_{2}\bigg) \;,
\end{align}
is the gauge potential, and 
\begin{align}\label{eq:B}
	\mathbf{B} &= \grad \times \mathbf{A}  \;,
\end{align}
is the corresponding magnetic field.  Physically, $\chi$ measures the linking number between the level sets (i.e. the preimages) of any two distinct points on $S^2$. However, it is worth noting that $\chi$ becomes trivial when an additional band is introduced into the relevant subspace (conduction or valence), so that these insulators are often described as having a delicate topology  \footnote{This is different from fragile topology where one is allowed to introduce additional valence bands while keeping the conduction band subspace fixed.\cite{Kennedy2016bott,Nelson2021}}.
\begin{figure*}[ht!]
	\centering
	\includegraphics[page=3,scale=1.3]{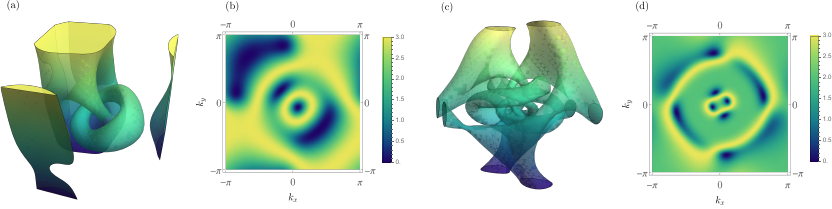}
	\caption{Isosurfaces and corresponding contour plots of the equal-time structure factor for fragile classical spin liquids (CSLs) derived from rational maps. (a) Three-dimensional isosurface of $S(\mathbf{k})$ for the Hopf map, Eq.~(\ref{eq:Hopfmap}).
		(b) Two dimensional contour plot of $S(k_x,k_y,0)$ for the Hopf map. (c) Three-dimensional isosurface of $S(\mathbf{k})$ for the Whitehead map, Eq.~(\ref{eq:Whiteheadmap}) with $p = 2, q=3$. (d)  Two-dimensional contour plot of  $S(k_x,k_y,0)$ for the same Whitehead map. The 3D isosurfaces directly visualize the knot or link topology encoded by the rational map, since the eigenvectors defining $S(\mathbf{k})$ span the plane perpendicular to the pseudo-spin field. Correspondingly, the 2D contour plots at $k_z = 0$ provide planar projections of these knotted or linked textures, offering insight into crossings and linking structure through their intersection patterns and concentric contours.}
	\label{fig:StructFact}
\end{figure*}

{\it Encoding complex topologies through rational maps\textbf{\----}} 
Rational maps provide a very natural means to encode various classes of knotted and linked structures. These maps are defined as the ratio of two complex polynomials, where the encoded topology can be viewed through it's level sets. 
Here we identify the map $\mathbf{z}$ with a rational map,
\begin{equation}\label{eq:rationalmap}
	\psi = \frac{z_{1}}{z_{2}} = \frac{N}{D}\;,
\end{equation}
where different choices of $N$ and $D$ can realize various kinds of topological insulators. In general, $N$ and $D$ can be thought of as arbitrary functions of $u_1$ and $u_2$, and their complex conjugates,
\begin{align}
	N = N(u_{1},u_{1}^{*},u_{2},u_{2}^{*}), D = D(u_{1},u_{1}^{*},u_{2},u_{2}^{*})\;.
\end{align}

{\it Hopf-Pontrjagin index of rational maps\textbf{\----}}
The Hopf map, for instance, can be described by the rational map
\begin{equation}\label{eq:Hopfmap}
	N = u_{1},\; D = u_{2}\;.
\end{equation}
This map was first considered in \cite{Moore2008}, where, for $m = -\frac{3}{2}$, the topological invariant was shown to be,
\begin{equation}\label{eq:Hopfmapinvariant}
	\chi = 1\;.
\end{equation} 

A generalization was later proposed in \cite{Deng2013} using the Whitehead map \cite{Whitehead1947expression}, defined as
\begin{equation}\label{eq:Whiteheadmap}
	N = u_{1}^{p},\; D = u_{2}^q\;,
\end{equation} 
where $p$ and $q$ are integers. In this scenario, the HP invariant becomes
\begin{eqnarray}
	\chi =\begin{cases}
		0, & \text{for } |m| > 3 \\
		pq, & \text{for } 1<|m|<3  \\
		-2pq &\text{for } |m| < 1\;.
	\end{cases}
\end{eqnarray}
When $p = q = 1$, we obtain the Hopf map Eq.~(\ref{eq:Hopfmap}).  

One can further extend these rational maps to encode torus knots and links via:
\begin{equation}\label{eq:Torusknotlinkmap}
	N = u_{1}^{\alpha} u_{2}^{\beta},\; D = u_{1}^{\gamma} + u_{2}^{\delta}\;,
\end{equation}
yielding the invariant
\begin{eqnarray}
	\chi =\begin{cases}
		0, & \text{for } |m| > 3 \\
		\alpha\delta + \beta\gamma, & \text{for } 1<|m|<3  \\
		-2(	\alpha\delta + \beta\gamma) &\text{for } |m| < 1\;.
	\end{cases}
\end{eqnarray}

The possible choices are not limited to torus knots and links. A broader family of rational maps, useful for encoding structures such as cable knots, fig-8 knot, Borromean rings, etc., is given by
\begin{equation}\label{eq:TuneTop}
	N = u_{1}^{\alpha}, D = D(u_{1},u_{1}^{*},u_{2})\;.
\end{equation}
where the topology is controlled by the choice of the polynomial $D$. For this class, the associated HP invariant is
\begin{eqnarray}
	\chi =\begin{cases}
		0, & \text{for } |m| > 3 \\
		\alpha \deg(D)_{u_{2}}, & \text{for } 1<|m|<3  \\
		-2(	\alpha \deg(D)_{u_{2}}) &\text{for } |m| < 1\;,
	\end{cases}
\end{eqnarray}
where $ \deg(D)_{u_{2}}$ corresponds to the highest power of $u_{2}$. 
For example, the complex polynomial
\begin{equation}\label{eq:C2332cableknot}
	D = u_2^{4} - 2 u_{1}^{3}u_{2}^{2} - 2i u_{1}^{3}u_{2} + u_{1}^6 +\frac{1}{4}u_{1}^3 \;,
\end{equation}
 encodes the cable-knot, $C^{2,3}_{3,2}$ \cite{Jennings2015cabling}, as visualized by its zero set ($D=0$), shown in [Fig.~\ref{fig:ZeroSetD}(c)]. 
 
Another versatile family of polynomials is the lemniscate family \cite{Bode2017knotted}, parameterized by three positive integers $(s,l,r)$. Specific choices within this family represent a wide range of knots and links: for example, $(s,r,l=1)$ yield torus knots, whereas $(s=3,r=3,l=2)$ corresponds to the Borromean rings. These polynomials have the form
\begin{align}\label{eq:Lemniscateknot}
	D = \prod_{m = 1}^{s} \bigg[u_{2} -& \frac{a}{2}\bigg(u_1^{r/s} e^{2\pi i m/s} + (u_1^{*})^{r/s} e^{-2\pi i m/s}\bigg)  \nonumber\\
	-& \frac{b}{2l}\bigg( u_1^{rl/s} e^{2\pi i ml/s} + (u_1^{*})^{rl/s} e^{-2\pi i ml/s}  \bigg) \bigg]\;,
\end{align}
with parameters $a = b = 1$. Once again, we can visualize these through their zero set. 

However, the polynomial associated with a given knot or link is not unique. For example, the complex polynomial \footnote{Note this corresponds to $(s = 3,r=2,l=2)$ in Eq.~(\ref{eq:Lemniscateknot}).}
\begin{align}\label{eq:fig8knot}
	D = & \;64 u_{2}^{3} - 12 u_{2} ( 3 + 2 u_{1}^2 - 2 u_{1}^{*2} )\nonumber \\
	& + (14 u_{1}^2 + 14 u_{1}^{*2}  + u_{1} ^{4} - u_{1}^{*4})\;,
\end{align}
encodes a figure-8 knot, [Fig.~\ref{fig:ZeroSetD}(b)], but the same knot is also represented by Rudolph’s polynomial \cite{Rudolph1987isolated}
\begin{equation}\label{eq:Rudolphmap}
	D = u_{2}^{3} -3 u_{2}(u_1u_{1}^*)^{2}(1+u_1^2 + u_1^{*2}) - 2 (u_1^2+u_1^{*2})\;.
\end{equation}

The approach, based on rational maps, thus provides a unified framework that not only consolidates previous methods but also naturally allows further generalization and connections across different fields, as discussed below.

{\it Magnetic fields are level curves of rational maps\textbf{\----}}
Field lines of electromagnetic fields constructed from rational maps are intricately linked \cite{Arrayas2017knots,Kedia2016}. Similarly, the magnetic field lines in Hopf insulators reflect the underlying topology encoded by these rational maps. To illustrate this, we rewrite the magnetic field making use of Eqs.~(\ref{eq:AandB}) and (\ref{eq:B}) as
\begin{equation}\label{eq:Magfiledaslevelcurve}
	\mathbf{B} = \frac{i  \mathbf{\nabla}\psi^{*} \times  \mathbf{\nabla}\psi}{(1+ \psi\psi^{*})^{2}}= \frac{-\mathbf{\nabla} \times \textsf{Im}(\psi^{*}\mathbf{\nabla}\psi)}{(1+ \psi\psi^{*})^{2}}\;.
\end{equation}
 The appearance of the curl, $\mathbf{\nabla} \times \textsf{Im}(\psi^{*}\mathbf{\nabla}\psi)$, implies that $\mathbf{B}$ is tangent to the level sets of $\psi$. This correspondence is demonstrated
  in [Fig.~(\ref{fig:StreamlineRationalMap})] for the Hopf map, Eq.~(\ref{eq:Hopfmap}), and the Whitehead map, Eq.~(\ref{eq:Whiteheadmap}) with~$p = 2, q =3$. where the level sets are higlighted in red and blue, resepectively. 
  \begin{figure}[ht!]
	\includegraphics[page=4,scale=1.2]{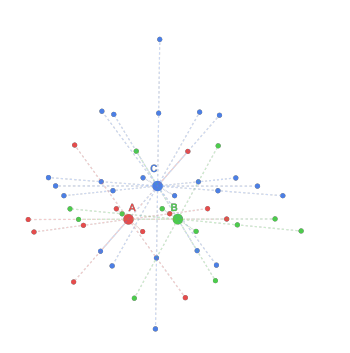}
  	\caption{Real-space constrainer representing the three-sublattice geometry. The sublattice sites are drawn in red, green, and blue (labeled $\alpha = A,B$ and $C$) and are positioned at $\mathbf{a}_{\alpha,j}$, Eq.~(\ref{eq:absolposit}). The bigger circles correspond to the respective sublattice origin $\mathbf{r_{\alpha}}$, Eq.~(\ref{eq:sublattorigin}). This plot was obtained by inverting the reciprocal‐space constrainer $\mathbf{\hat{n}(k)}$ contructed from Hopf map, Eq.~(\ref{eq:Hopfmap}). }
  	\label{fig:HopfConstrainerReal}
  \end{figure}

\begin{figure*}[t!]
	\centering
	\includegraphics[page=5,scale=1.2]{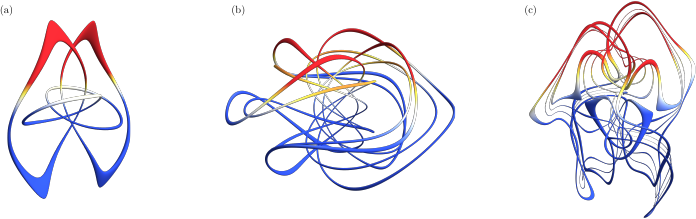}	
	\caption{Demonstrating different types of nodal lines topologies. (a) Nodal lines for Hopf map, Eq.~(\ref{eq:Hopfmap}). (b) Nodal lines for Whitehead map, Eq.~(\ref{eq:Whiteheadmap}) with $p = 2, q=2$.
		(c) Nodal lines for the map in Eq.~\ref{eq:TuneTop} with $\alpha = 1$ and $D$ given by Eq.~(\ref{eq:Rudolphmap}). These plots were obtained for $m=-2.3$ and $\lambda = 0.8$.}
	\label{fig:NodalLines}
\end{figure*}

{\it Isosurfaces of Fragile Classical Spin Liquids \textbf{\----}}
Employing the pseudo‑spin field $\mathbf{\hat{n}(k)}$, we construct three-dimensional gapped CSLs, characterized by the HP invariant. Following the terminology of \cite{Han2024}, these systems fall under the fragile topological CSL class. The CSL Hamiltonian in momentum space is given by
\begin{equation}
	\mathcal{H}_{\sf CSL} = \frac{1}{2}\sum_{\mathbf{k}}\sum_{a,b = 1}^{N = 3} \tilde{S}_{a}(-\mathbf{k})\bigg[\mathcal{H}_{\sf ps}(\mathbf{k})\bigg]_{ab}\tilde{S}_{b}(\mathbf{k})\;,
\end{equation}
where, $\tilde{S}_{a}$ is the Fourier transform of spin field $S_{a}$, and $a,b$ label the sublattice sites. The pseudo-spin Hamiltonian is defined as
\begin{equation}
	\mathcal{H}_{\sf ps}(\mathbf{k}) =\mathbf{\hat{n}(k)} \otimes \mathbf{\hat{n}(k)}\;.
\end{equation}

Here the pseudo-spin field functions as a constrainer in reciprocal space.
For further details on the constrainer, please refer to the End Matter.

By construction, the pseudo‑spin Hamiltonian, $\mathcal{H}_{\sf ps}(\mathbf{k})$ has a gapped spectrum, 
characterized by the HP invariant of the chosen rational map. 
For example, employing the Hopf map, Eq.~(\ref{eq:Hopfmap}) results in a system classified by its corresponding HP invariant, Eq.~(\ref{eq:Hopfmapinvariant}), albeit with an intricate real-space constrainer [Fig.~(\ref{fig:HopfConstrainerReal})] (refer to End Matter). 

To experimentally probe the embedded topological structure, we look at the  equal time structure factor 
\begin{equation}
	S(\mathbf{k}) =  \sum_{i : \omega_i = 0} \bigg|\sum_{a = 1}^{3} v_{i}^{a}(\mathbf{k}) \bigg|^{2}\;.
\end{equation}
Here, $\omega_i$ denote the eigenvalues of $\mathcal{H}_{\sf ps}(\mathbf{k})$, and the eigenvectors $v_{i}(\mathbf{k})$ span the degenerate subspace corresponding to the zero eigenvalues. The resulting isosurfaces of $S(\mathbf{k})$ directly capture the nontrivial linking structure of the pseudo-spin field's level set, as explicitly demonstrated in [Fig.~(\ref{fig:StructFact})]. Additionally, the two-dimensional contour plots, such as the $S(k_x,k_y,0)$ slice, serve as planar projections clearly revealing the crossing and linking patterns underlying these topological textures. The structure factor thus provides a practical and measurable fingerprint of the intricate topology inherent to our CSL construction.

{\it Nodal lines of Dispersive Hopf insulators\textbf{\----}}
Level sets of rational maps can also be imprinted onto the nodal lines of dispersive Hopf insulators \cite{Jankowski2024}. These insulators are described by the Hamiltonian
\begin{equation}\label{eq:DispHopf}
	\mathcal{H}_{\sf disp}(\mathbf{k}) = 2 \mathbf{\hat{n}(k)} \otimes  \mathbf{\hat{n}(k)}- I_{3} + \lambda  \operatorname{diag}\{-1,0,1\}\;,
\end{equation}
where $\lambda$ is a band-gap parameter (chosen as $\lambda = 0.8$ here). In this framework, the nodal lines defined by the degeneracy condition,
\begin{equation}
	E_1(\mathbf{k}) = E_2(\mathbf{k})\;,
\end{equation}
for the lowest two eigenvalues, exhibit the same topological structure as the level sets of the rational map \cite{Jankowski2024}. For example, when the Hopf map, Eq.~(\ref{eq:Hopfmap}) is employed, the nodal lines form a Hopf link, as illustrated in [Fig.(\ref{fig:NodalLines})]. By choosing other rational maps, one can similarly generate topologies such as torus knot,  [Fig.~(\ref{fig:NodalLines})], figure-8 knot, [Fig.~(\ref{fig:NodalLines})] etc. The key principle is the same: the chosen map enforces the relevant topological structure.

{\it Zero sets of Topological semimetals, and Non-Hermitian metals\textbf{\----}}
We already showed (see Fig. \ref{fig:ZeroSetD}) how the zero sets of complex polynomials can encode nontrivial knot or link structures. 
We leverage this idea to model more exotic kinds of topological semimetals generalizing the construction of \cite{Ezawa2017,Yan2017,Chen2017}. In particular, we consider the $\mathcal{PT}$-symmetric Hamiltonian
\begin{equation}
	\mathcal{H}_{\textsf{TS}}(\mathbf{k})  = a_{1}(\mathbf{k}) \sigma^{x} + a_{3}(\mathbf{k})\sigma^{z}\;.
\end{equation}
where
\begin{equation}
	a_{1} = \textsf{Re}(D),\; a_3 =  \textsf{Im}(D)
\end{equation}
The energy bands, $E_{\pm} = \pm \sqrt{a_1^2 + a_3^2}$, vanish precisely at points where $D=0$. By choosing different complex polynomials $D$, one can embed various knots or links into these zero sets. In [Fig.~(\ref{fig:ZeroSetD})], the blue curves illustrate several such examples for different choices of $D$, indicating where $E_{\pm}$ goes to zero. 

This same idea can be extended to non-Hermitian systems, allowing the construction of knotted or linked metals beyond the torus-based class \cite{Yan2017}.  Consider the non-Hermitian Hamiltonian
\begin{equation}\label{eq:NHHam}
	\mathcal{H}_{\textsf{NH}}(\mathbf{k})  =  \mathbf{d_{R}}(\mathbf{k}) \cdot  \boldsymbol{\sigma} + \mathbf{d_{I}}(\mathbf{k})\cdot  \boldsymbol{\sigma}\;,
\end{equation}
where
\begin{equation}
	\mathbf{d_{R}} = (a_1- \Lambda, \Lambda,0),\; \mathbf{d_{I}} = (0,a_{3}, -\sqrt{2}\Lambda) \;.
\end{equation}

For large $\Lambda$, the zero sets of $D$  in this framework corresponds to the exceptional points of $\mathcal{H}_{\textsf{NH}}$ \cite{Carl2019}. For instance, the torus class can be encoded using $D$ in Eq.~(\ref{eq:Torusknotlinkmap}). For the lemniscate family, Eq.~(\ref{eq:Lemniscateknot}) can be used. For cable knot  $C^{2,3}_{3,2}$, Eq.~(\ref{eq:C2332cableknot}) and so on. 

{\it Summary\textbf{\----}}
In this work, we have developed a unified method to systematically embed complex knots and links into a diverse range of topological systems, including topological insulators, classical spin liquids, topological semimetals, and non-Hermitian metals. By utilizing rational maps and the level sets of complex polynomials, our construction bypasses the need for separate parameterizations, directly translating knot and link topologies onto physical models. This approach allowed us to explicitly construct new models exhibiting intricate topological textures in each of the considered platforms.
Specifically, we demonstrated how Hopf insulators naturally embody aspects of topological electromagnetism through the emergent magnetic field lines.
Moreover, for the newly introduced fragile classical spin liquids, we showed that the experimentally accessible equal-time structure factor directly reflects the encoded level-set topology. Looking forward, this method suggests intriguing possibilities for discovering similarly rich textures in Floquet~\cite{Cayssol2013floquet}, crystalline~\cite{Fu2011topological}, and higher-order topological insulators~\cite{Schindler2018higher}, paving the way for designing novel topological phases and revealing deeper connections among diverse physical platforms.

{\it Acknowledgments\textbf{\----}}
This work is supported by the Theory of Quantum Matter Unit of the Okinawa Institute of Science and Technology Graduate University (OIST). We would like to thank Nic Shannon, Tokuro Shimokawa, Yoshi Kamiya, Jan Willem Dalhuisen, Pranay Patil, and Jiahui Bao for their valuable comments and feedback on the manuscript. We also thank Arthur Morris for highlighting his work \cite{Jankowski2024} during his visit at OIST.

\bibliography{document.bib}
\appendix
\section*{End Matter}
In the constrainer formalism for classical spin liquids \cite{Han2024}, the local ground state constraint is enforced by a Hamiltonian in real space
\begin{equation}
	\mathcal{H} = \sum_{\mathbf{R}}[\mathcal{C}(\mathbf{R})]^{2}
\end{equation}
where the sum runs over all unit cells labeled by the Bravais lattice vector $\mathbf{R}$. Minimization of $\mathcal{H}$ forces the local constraint 
\begin{equation}
	\mathcal{C}(\mathbf{R}) = 0 \;\forall \text{ unit cells.}
\end{equation}
The local constrainer $\mathcal{C}(\mathbf{R})$ is written as
\begin{equation}
	\mathcal{C}(\mathbf{R}) = \sum_{\mathbf{r}}\mathbf{S}(\mathbf{r})\cdot \mathbf{C}(\mathbf{R},\mathbf{r})
\end{equation}
where $\mathbf{S}(\mathbf{r})$ is the vector array of spins (with components corresponding to the different sublattice sites) and $\mathbf{C}(\mathbf{R},\mathbf{r})$ is a vector that encodes the weights by which these spins contribute to the constrainer. 

 For a system with $N$ degrees of freedom per unit cell, the constraints can be expressed as an $N$-component vector, $\mathbf{C}(\mathbf{R},\mathbf{r})$.  In our case, $N=3$, corresponding to three sublattice sites. We define the constrainer at the origin (i.e. $\mathbf{R} = \mathbf{0}$) in a vector form that explicitly encodes how spins from different sublattices are summed:
\begin{equation}
	\mathbf{C}(0,\mathbf{r}) = \begin{pmatrix} \sum_{j\in \mathrm{A}} c_{A,j}\,\delta_{\mathbf{r},\mathbf{a}_{A,j}} \\[1mm] \sum_{j\in \mathrm{B}} c_{B,j}\,\delta_{\mathbf{r},\mathbf{a}_{B,j}} \\[1mm] \sum_{j\in \mathrm{C}} c_{C,j}\,\delta_{\mathbf{r},\mathbf{a}_{C,j}} \end{pmatrix}\;.
\end{equation}
Here, $\mathbf{r}$ runs over all lattice sites, and $\delta_{\mathbf{r},\mathbf{a}_{\alpha,j}}$ picks out the spins at the positions $\mathbf{a}_{\alpha,j}$. We decompose these positions as
\begin{equation}
	\mathbf{a}_{\alpha,j} = \mathbf{R} + \mathbf{r}_{\alpha} + \delta\mathbf{a}_{\alpha,j}
\end{equation}
where $\mathbf{R}$ is the Bravais lattice vector (with a simple cubic structure in our model), $ \mathbf{r}_{\alpha} $ is the chosen origin for sublattice, and $\delta\mathbf{a}_{\alpha,j}$ is the displacement (or internal coordinate) of the $j$th site within that sublattice.

\begin{figure*}[t!]
	\centering
	\includegraphics[page=6,scale=1.3]{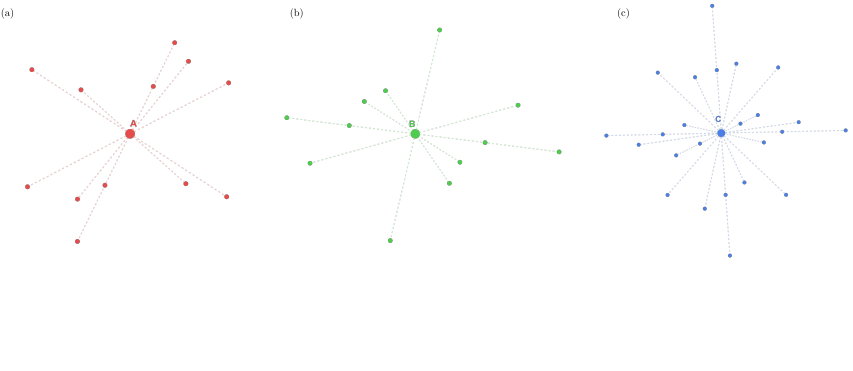}
	\vspace{-2.3cm}
	\caption{Highlighting the different sublattices for the momentum space constrainer obtained from the Hopf map. (a) $A$ sublattice sites. (b) $B$ sublattice sites. (c) $C$ sublattice sites. The bigger circles correspond to the respective sublattice origin $\mathbf{r_{\alpha}}$, Eq.~(\ref{eq:sublattorigin}).}
	\label{fig:Sublattices}
\end{figure*}

To obtain the real‐space sublattice sites [Fig.~(\ref{fig:HopfConstrainerReal})] from the pseudo-spin field $\mathbf{\hat{n}(k)}$ (which is the Fourier transform of $\mathbf{C}(0,\mathbf{r})$) we perform an inverse Fourier transform. Each component of $\mathbf{\hat{n}(k)}$, corresponds to a sublattice $(\alpha = A,B,C)$, from which we extract the displacement vectors $\delta\mathbf{a}_{\alpha,j}$.  
We then choose sublattice origins:
\begin{equation}\label{eq:sublattorigin}
	\mathbf{r}_A = (0,0,0),\quad 
	\mathbf{r}_B = \Bigl(\tfrac{1}{2},\tfrac{1}{2},0\Bigr),\quad 
	\mathbf{r}_C = \Bigl(\tfrac{1}{2},0,\tfrac{1}{2}\Bigr).
\end{equation}
The absolute positions where spins reside are then given by
\begin{equation}\label{eq:absolposit}
	\mathbf{a}_{\alpha,j} = \mathbf{r}_\alpha + \delta\mathbf{a}_{\alpha,j}.
\end{equation}
These are plotted in red, green, and blue for sublattices A, B, and C, respectively. See Fig.~\ref{fig:HopfConstrainerReal} and Fig.~\ref{fig:Sublattices} for the pseudo spin field corresponding to the Hopf map.

\end{document}